\documentclass[showpacs,preprintnumbers,amsmath,amssymb,nofootinbib]{revtex4}
\usepackage{graphicx}
\usepackage{color, soul}
\usepackage{amsmath}
\usepackage{amsfonts}
\input epsf
\usepackage[pdftex,
            pdfauthor={A. V. Yurov, V. A. Yurov},
            pdftitle={The d-Dimensional Cosmological Constant and the Holographic Horizons},
            pdfsubject={},
            pdfkeywords={Friedmann equations; higher dimensions; Sturm--Liouville problem; fractal event horizons},
            pdfproducer={Latex with hyperref},
            pdfcreator={pdflatex}]{hyperref}
\hypersetup{colorlinks=true, citecolor=blue, linkcolor=red}
\usepackage{enumerate}
\usepackage{scrextend}
\addtokomafont{labelinglabel}{\sffamily}

\newcommand{\beq}{\begin{equation}}
\newcommand{\beqn}{\begin{equation*}}
\newcommand{\enq}{\end{equation}}
\newcommand{\enqn}{\end{equation*}}
\newcommand{\eb}{{\rm e}}

\newcommand{\N}{{\mathbb N}}

\renewcommand{\l}{\lambda}

\newtheorem{Hypothesis}{\textsc{Hypothesis}}

\begin{document}
\title{The d-Dimensional Cosmological Constant and the Holographic Horizons}
\author{A.V. Yurov}%
\email{AIUrov@kantiana.ru}
\author{V.A. Yurov}%
\email{vayt37@gmail.com}
\affiliation{I. Kant Baltic Federal University, Department of Physics, Mathematics and IT, Al. Nevsky str. 14, Kaliningrad
236041, Russia}

\begin{abstract}
This article is dedicated to establishing a novel approach to the cosmological constant, in which it is treated as an eigenvalue of a certain Sturm--Liouville problem. The key to this approach lies in the proper formulation of physically relevant boundary conditions. Our suggestion in this regard is to utilize the ``holographic boundary condition'', under which the cosmological horizon can only bear a natural (i.e., non-fractional) number of bits of information. Under this framework, we study the general d-dimensional problem and derive the general formula for the discrete spectrum of a positive energy density of vacuum. For the particular case of two dimensions, the resultant problem can be analytically solved in the degenerate hypergeometric functions, so it is possible to define explicitly a self-action potential, which determines the fields of matter in the model. We conclude the article by taking a look at the d-dimensional model of a fractal horizon, where the Bekenstein's formula for the entropy gets replaced by the Barrow entropy. This gives us a chance to discuss a recently realized problem of possible existence of naked singularities in the $D\neq 3$ models.
\end{abstract}

\maketitle
\section{Introduction} \label{sec:Intro}
\allowdisplaybreaks

It is a clear and undeniable fact that we live in a Fridemann universe that has exactly $(1+3)$ dimensions. It is also a fact that the string theory authoritatively asserts the existence of no less than 11 dimensions. These seemingly mutually exclusive claims can be reconciled if we accept that in six out of seven ``superfluous'' dimension in the observable universe are compactified into a Calabi--Yau manifold, while the remaining dimension, albeit macroscopic, is unreachable to us (except gravitationally), beings locked up on a $(1+3)$-dimensional brane. Naturally, the string theory also predicts an existence of a multitude of other pocket universes with the characteristics wildly different from ours. Together they form what is called a {string landscape} -- an infinite collection of pocket universes characterized by different types of compactification. And while the total number of possible compactifications is finite, it is also staggeringly huge, reaching about $10^{500}$. This fact provides us with an unequivocal evidence of simple yet important fact: the overwhelming majority of the universes in the string landscape looks nothing like the observable universe. In other words, while we as observers can only exist in a $(1+3)$-dimensional subset of string landscape \cite{Barrow_Tipler}, the total share of this subset is infinitely small. And this naturally opens up a rather intriguing question: what kind of cosmological models might we expect in those more prevalent pocket universes?.. As we shall see, in order to answer this question will require some very interesting mathematical tools and theoretical implements. In particular, in this article we follow the idea which has originated in our earlier paper \cite{Sturm}, where we have proposed that the exact value of the vacuum energy might be treated as an eigenvalue of a certain Sturm--Liouville problem with some specific boundary conditions. We applying this idea for a $d$-dimensional Friedmann model by introducing the following boundary condition: the cosmological event horizon must be such that it contains a finite {natural} number of bits of information. This seemingly minor hypothesis (subsequently dubbed the {Main Hypothesis}) then ends up being a key to the entire discussion, and warrants a separate discussion which will thus be postponed until Section \ref{sec:Quantization}. Essentially, we are using a very specific form of a holographic Bekenstein \linebreak bound \cite{Bek-2} in order to find the correct boundary conditions  on the spectrum of permissable values of the vacuum energy. Note that although we do not use the holographic dark energy models (see \cite{HOLO-0,HOLO-1,HOLO-2}) it is quite possible that the boundary conditions we postulate in this paper can also be derived in the framework of those models.

{We emphasize once again that in this article we do not work with a holographic cosmological model; instead, we use the holographic bound, originally established for the horizons of the black holes, applying it for a completely different object: a dS horizon. More precisely, we propose a hypothesis that a de Sitter entropy behaves similarly to the Bekenstein-Hawking entropy of a black hole, which must be proportional to the surface area of the horizon. Although this statement is widely used by various authors, it is important to understand the fundamental difference between those two types of horizons: the dS horizon is necessarily observer-dependent, and the black hole horizon is not. This particular distinction immediately makes doubtful a simple straightforward transfer of the properties of the black hole horizons onto a de Sitter horizon. Nevertheless, we would argue that these two different types of the horizons shall indeed share their properties, thanks to the following remarkable fact: the gravitational radius (calculated according to the Schwarzschild formula) for the de Sitter region filled with a positive cosmological constant exactly coincides with the dS radius, and it happens regardless of the choice of $d$} (see Formula (\ref{RandR})). {In other words, the two different radiuses appear to be naturally intertwined with each other, so that whenever we are discussing the radius of the dS horizon, we might invoke the properties specific to the gravitational radius, such as the Bekenstein-Hawking constraint. Admittedly, this is not a rigorous proof, but it serves its purpose as a solid foundation for the subsequent calculations.}

Before we delve deeper into the problem, let us briefly discuss the structure of this paper. In Section \ref{sec:Black_hole_capacity} we will show how, using a very simple heuristic method, one can derive the general formula for the informational capacity of $d$-dimensional black holes, and then, by establishing the relationship between 1 bit of information recorded on the black hole's horizon with a unitary $d$-dimensional Plank surface area we find the exact formula for the latter. On our next step in Section \ref{sec:linearization} we introduce a useful formalism, based on the results of \cite{Tem} generalized for the $d$-dimensional case. This new formalism would be a huge aid in studying the dynamics of multidimensional Friedmann models, because it would allow us to effectively reduce the problem to a single linear Schr\"odinger equation, with the positive cosmological constant taking the guise of an eigenvalue of that equation. However, in order to properly define this eigenvalue a correct boundary condition must be introduced. This we do in Section \ref{sec:Quantization}, using the aforementioned Main Hypothesis: that the surface area of any physically feasible cosmological horizon shall contain a {natural} number of bits of information. Under this assumption the Friedmann model severely limits possible cosmological constants to a countable set parameterized by a single integer. And if the remaining fields of matter are modeled by a single minimally connected scalar field, the data we got would be sufficient for calculating the self-action potential. This will be aptly demonstrated in a simple case $d=2$ where all the calculations can be performed in the explicitly analytic manner with the aid of the hypergeometric function. In the next section, Section \ref{sec:fractals} we turn our attention to the generalization of  the ``Barrow entropy'' \cite{Bar-1,Bar-2} for $d$ dimension. In particular, we demonstrate that even in the Barrow framework where the structure of the horizon ends up being essentially fractal, this fractality changes our model but a little, and predicts no significant alterations in the set of permissible cosmological constants. Despite the fact that in the fractal case the vacuum energy's density depends not on one but on two natural numbers, their values are to so large that the second of them essentially ceases to influence the results in a meaningful manner. However, in conclusion (see Section \ref{sec:conclusion}) we discuss one possible exception of this rule: it is possible that for the cosmologies  with $d>3$ an extreme case of fractality must be taken into account, namely: an infinite horizon encompassing a finite volume. The properties of such a pocket universe would be most intriguing, however, it must be owned that Barrow himself deems such any model ``unphysical''  \cite{Bar-1}. Finally, in the Appendix \ref{sec:appendix} we demonstrate in the general case of a $d$-dimensional closed universe where the matter's equation of state is equal to $p/c^2=w\rho$ with $w$ = const, the resulting Friedmann equations can be completely integrated for any $w$.

\section{The Informational Capacity of a \boldmath{$d$}-Dimensional Black Hole}\label{sec:Black_hole_capacity}

While the concept of informational capacity of a black hole still manages to astonish the mind as something alien to the common sense, the actual value of this capacity can be derived without much ado in a process that has even been featured in the popular literature---see, for example, the beautiful book \cite{Susk}.

Let us go over this derivation in the case of $d$ dimension. For this end, consider a Schwarzschild solution describing a $d$-dimensional black hole of mass $m$ and gravitational radius $R_{(d)}$ \cite{1963}:
\begin{equation}
ds^2=-c^2\left(1- \left(R_{(d)}/r\right)^{d-2}     \right)dt^2+\frac{dr^2}{1-\left(R_{(d)}/r\right)^{d-2}}+r^2d\Omega^2,
\label{1.1}
\end{equation}
where $\Omega$ denotes the surface of a $(d-1)$-dimensional unit sphere, whose entire surface area $\Omega_{d-1}=2\pi^{d/2}/\Gamma(d/2)$ (here $\Gamma(x)$ is the Gamma function defined as $\Gamma(x)=\int_0^\infty t^{x-1} \eb^{-t} dt$.), the gravitational radius $R_{(d)}$ has the form:
\begin{equation}
R_{(d)}=\left[\frac{16\pi G_{(d)}m}{(d-1) \ c^2 \ \Omega_{d-1}}\right]^{1/(d-2)},
\label{1.2}
\end{equation}
and $G_{(d)}$ denotes the $d$-dimensional gravitational constant. Note, that in the international system of units the dimension of $G_{(d)}$ would be $\left[\rm {meter}^d/ (\rm {kilogram}\cdot {\rm second}^2)\right]$. Once we know this, it is then straightforward to check that the $d$-dimensional Plank length has the following relationship with $G_{(d)}$, Plank constant $\hbar$ and the speed of light $c$:
\begin{equation}
L_{(d)PL}=\alpha\left(\frac{\hbar G_{(d)}}{c^3}\right)^{1/(d-1)},
\label{1.3}
\end{equation}
where $\alpha$ is a dimensionless proportionality coefficient which will be precisely defined a bit later.

Now we need to know how many bits of information can be recorded on the $(d-1)$-dimensional surface of the black hole in consideration. In order to figure this out, we'll need two crucial pieces of information. First, we know that the total surface area of this black hole is $A_{(d)}=\Omega_{d-1}R_{(d)}^{d-1}$. Second, if a photon of wave length $\lambda=2R_{(d)}$ gets swallowed by this black hole, its horizon gains exactly one bit of information (the photons with larger wave lengths will simply bypass the black hole without being consumed, see \cite{Susk}). During this process the mass of our black hole increases by

$$
m_{\gamma}=\frac{\pi\hbar}{cR_{(d)}},
$$
while its surface area grow by
$$
\Delta A_{(d)}=\Omega_{d-1}\left({\tilde R}_{(d)}^{d-1}-R_{(d)}^{d-1}\right),
$$
where ${\tilde R}_{(d)}$ can be found from (\ref{1.2}) by changing $m\to m+m_{\gamma}$. Since $m_{\gamma}\ll m$, we can expand the resulting expression into the Taylor series over parameter $m_{\gamma}/ m$. Neglecting all the higher order terms yields the following formula:
 \begin{equation}
\Delta A_{(d)}=\frac{16\pi^2}{d-2}\left(\frac{L_{(d)PL}}{\alpha}\right)^{d-1}.
 \label{1.4}
 \end{equation}

According to (\ref{1.4}), by defining the hitherto unknown parameter $\alpha$ as
 \begin{equation}
\alpha=\left(\frac{16\pi^2}{(d-2)\Omega_{d-1}}\right)^{1/(d-1)},
 \label{1.5}
 \end{equation}
we will ensure that one bit of information will correspond to exactly one $d$-dimensional Plank area $A_{(d)PL}=\Omega_{d-1}L_{(d)PL}^{d-1}$. We would like to emphasize that the appearance if a factor $r^{d-1}$ is a direct outcome of the assumption that the increment $\Delta A_{(d)}$ depends not on the gravitational radius $R_{(d)}$, but is instead a universal (dimensional) constant.

\section{The Linearization of \boldmath{$d$}-Dimensional Friedmann Equations}\label{sec:linearization}

In this paper for simplicity's sake we will mostly stick to Friedmann models with flat geometries (the closed Friedmann models with radiation dominance have previously been studied in \cite{Ver}, and the more general models including a multicomponent coupled fluid were investigated in \cite{Ver-Od}); however, we will take a look at some of the closed models in the Appendix \ref{sec:appendix}.

Now, let $H$ denote the Hubble parameter, $\rho$ -- density, and let $p$ stand for pressure. Then the dynamics of the flat $d$-dimensional isotropic and homogeneous universe will abide by the following Friedmann--Robertson--Walker--Lema\^itre equations (see \cite{Ver}:
\begin{equation}
H^2=\frac{16\pi G_{(d)}\rho}{d(d-1)},
\label{2.1}
\end{equation}
\begin{equation}
{\dot H}=-\frac{4\pi G_{(d)}}{d-2}\left(\rho+\frac{p}{c^2}\right).
\label{2.2}
\end{equation}

If the universe is filled primarily with radiation, due to the conformal symmetry the energy-momentum tensor ends up being traceless. This means that the equation of state reduces to a simple relationship $p=c^2\rho/d$. The model shall also acknowledge a presence of a positive vacuum energy (i.e., of cosmological constant $\Lambda$) with density $\rho_{_{\Lambda}}={\rm const}>0$. The remaining fields will be modeled by a minimally coupled scalar field $\phi$ with the potential $V(\phi)$. Tying it all together into one equation produces the following:
\begin{equation}
\frac{{\ddot a}}{a}=-\frac{8\pi G_{(d)}(d-2)}{d(d-1)}\left(\rho+\frac{p\, d}{c^2(d-2)}\right),
\label{2.3}
\end{equation}
where for convenience the $\rho_{_{\Lambda}}$ and $\phi$ contributions have been integrated into the general expressions for density and pressure
\begin{equation}
\rho=\frac{{\dot\phi}^2}{2}+V(\phi)+\rho_{_{\Lambda}},\qquad
p=\frac{{\dot\phi}^2}{2}-V(\phi)-\rho_{_{\Lambda}},
\label{2.4}
\end{equation}
and so is the factor $c^2$, which has been included into a redefined $p$. The Equations (\ref{2.1}) \linebreak and (\ref{2.2}) taken together produce a continuity equation
$$
{\dot\rho}=-H\left(\rho+p\right)d,
$$
and from it the field equation in the Friedmann metrics can be easily derived. Here it is:
\begin{equation}
{\ddot\phi}+H{\dot\phi}d+\frac{dV(\phi)}{d\phi}.
\label{2.5}
\end{equation}

As a next step, let us introduce an arbitrary real number $n$, and a new function $\psi_n=a^n$. Using (\ref{2.1})--(\ref{2.3}) it is a short work to prove that $\psi_n$ must satisfy the following Schr\"odinger equation:
\begin{equation}
\frac{{\ddot\psi}_n}{\psi_n}=-\frac{8\pi n G_{(d)}}{d(d-1)}\left((d-n){\dot\phi}^2-2n(V(\phi)+\rho_{_{\Lambda}})\right).
\label{2.6}
\end{equation}

Note, that while we can choose different values of $n$, if we choose $n=d$ the equation simplifies even further, turning into
\begin{equation}
{\ddot\psi}=\left(u(t)-E\right)\psi,
\label{2.7}
\end{equation}
where $\psi_d=\psi$ and
\begin{equation}
u(t)=\frac{16\pi d G_{(d)}V(\phi(t))}{d-1},\qquad  E=-\frac{16\pi dG_{(d)}\rho_{_{\Lambda}}}{d-1}.
\label{2.8}
\end{equation}

In this way we have constructed a $d$-dimensional generalization of the results of Chervon. What should be the next logical step? Since we are already dealing with the Schr\"odinger Equation (\ref{2.8}), it would be natural to treat it as a full-fledged Sturm--Liouville problem for the eigenvalue $E$. However, in order to do that, we have to introduce a boundary condition? It has be reasonable and follow logically from the most fundamental properties of our model. We have a suggestion on this regard, and it it this: the boundary condition necessarily arises when we count the total amount of information that can fit on the $d$-dimensional surface area of the event horizon. Let us discuss this in detail in the next section.

\section{``Quantizing'' the $ \rho_{_{ \Lambda}}$}\label{sec:Quantization}

Let's take a look at a flat universe filled with both the fields of matter and the vacuum energy. During the later parts of evolution of such a universe the fields of matter's contribution will gradually vanish and the cosmological dynamics gets determined by the cosmological constant. The asymptotic scale factor during this phase take a form of $a=\exp (H_{(d)}t)$, where
\begin{equation}
H_{(d)}=4\sqrt{\frac{\pi G_{(d)}\rho_{_{\Lambda}}}{d(d-1)}}.
\label{3.1}
\end{equation}

Such a dynamics generates an event horizon of radius $R_{(d)H}$ equal to
\begin{equation}
R_{(d)H}=c a(t)\int_t^{+\infty}\frac{dt'}{a(t')}=\frac{c}{H_{(d)}},
\label{3.2}
\end{equation}
{(note that if the lower limit is chosen to be $t=0$ we will instead get the radius of the particle horizon)}.

{Now it is easy to show that the gravitational radius corresponding to the ``vacuum mass'' confined within the dS horizon is exactly equal to the radius of the cosmological event horizon, i.e.}
\begin{equation}
R_{(d)H}=R_{(d)}.
\label{RandR}
\end{equation}

{In fact, the  ``vacuum mass'' is}
\begin{equation}
m=\frac{\Omega_{d-1} \rho_{_{\Lambda}} R_{(d)H}^d}{d}.
\label{mm}
\end{equation}

{We can then express the density $\rho_{_{\Lambda}}$ in terms of the radius $R_{(d)H}$  using} (\ref{3.1}) and (\ref{3.2}),  substitute it into the  (\ref{mm}) and then calculate the gravitational radius by the Formula (\ref{1.2}). {It is easy to check that the result will be the Expression} (\ref{RandR}).

Now the main hypothesis.
\begin{Hypothesis}
{The $d$-dimensional ``surface'' of cosmological horizon must be such as to contain a natural number of Plank areas, thus recording a natural number of bits of information. In other words, using (\ref{1.5}), our Hypothesis  implies that}
\begin{equation}
\left(\frac{R_{(d)H}}{L_{(d)PL}}\right)^{d-1}=N\in \N.
\label{inN}
\end{equation}
\end{Hypothesis}
{Before moving on, we should make a few comments. First, while the expression} (\ref{inN}) {share certain similarity with the formulas arising in the loop quantum gravity, it has little to do with them. Indeed, the loopback approach does demonstrate a discreteness of the area and the volume} \cite{Smolin}, {but their formulas are expressed in terms of the ``colors'' of the links adjacent to the corresponding nodes of the spin network and they differ from the Formula} (\ref{inN}) (see, however, \cite{Ashtekar}). {Secondly, although at first glance the Equation} (\ref{inN}) {might look too strong ``to believe'', it expresses a very simple statement: the smallest physically meaningful $d$-dimensional area is a $d$-dimensional Plank area} $A_{(d)PL}=\Omega_{d-1}L_{(d)PL}^{d-1}$. {Due to this, any real $d$-dimensional area must, obviously, be a multiple of a $d$-dimensional Planck area, which is what the Equation} (\ref{inN}) {expresses. This reasoning is rather similar to the idea of Bekenstein and Mukhanov that the event horizon of a black hole shall be quantizable in integers} \cite{Bekenstein,Mukhanov,Mukhanov_Bekenstein}. {Thirdly, apparently a condition akin to} (\ref{inN}) {is necessary for the quantum theory of gravity in de Sitter space to make sense at all. The point here is that the quantum Hilbert space in the de Sitter space is finite-dimensional and therefore the Einstein Lagrangian with a positive $\Lambda$-term cannot be quantized in a consistent manner, and it must be derived from a more fundamental theory that would determine the possible values of} $G_{(d)}\rho_{_{\Lambda}}^{(d-2)/d}$ \cite{Witten}, {because the dimension of  the quantum Hilbert space must  be  a nontrivial function of}
 $G_{(d)}\rho_{_{\Lambda}}^{(d-2)/d}$ {and at the same time an integer. It's doable if the possible values of} $\rho_{_{\Lambda}}$  {are discrete, that is, parameterized with an integer (or integers). We call this a ''quantization'' of the vacuum energy. The Equation} (\ref{inN}) {is but a straightforward realization of this idea. In fact}
after a few straightforward calculations this formula yields the following condition for ``quantization'' of the vacuum energy density in a $d$-dimensional space:
\begin{equation}
\rho_{_{\Lambda}}=\frac{d(d-1)  N^{2/(1-d)}}{16\pi G_{(d)}\tau_{(d)PL}^2},
\label{3.3}
\end{equation}
where $\tau_{(d)PL}=L_{(d)PL}/c$. If we substitute (\ref{3.3}) into (\ref{2.7}), the result would be the Schr\"odinger equation with the ``energy'' $E$ defined as \begin{equation}
E=-\frac{n^2}{2\tau_{(d)PL}^2 N^{2/(d-1)}}.
\label{3.5}
\end{equation}

Notably, a rate of change of the potential $u(t)$ in the aforementioned Schr\"odinger equation satisfy the following beautiful formula:
$$
{\dot u}=\frac{16\pi G_{(d)}n{\dot\phi}}{d-1}\left((d-n)H{\dot\phi}+\frac{dV}{d\phi}\right).
$$

In order to demonstrate the benefits of this method, let us consider a simple two-dimensional model (remember, that from a standpoint of string landscape, there ought to be many more versions of two-dimensional pocket universes than those with $d=3$; after all, there are more ways to compactify seven extra dimensions than six!)

When $d=2$, the ``energy condition'' (\ref{3.5}) produces a well-known ``Coulomb'' spectrum
\begin{equation}
E=-\frac{n^2}{2\tau_{(2)PL}^2 N^2}.
\label{quli}
\end{equation}

The potential $u=u(t)$ from the Schr\"odinger Equation (\ref{2.7}) turns into
$$
u(z)=\frac{4\kappa^2
l(l-1)}{z^2}-\frac{2\sqrt{2}\kappa |n|}{\tau_{(2)PL}^2 z},
$$
where $E=-\kappa^2$, $z=2\kappa t$, $l=1,2,...$, and we took an absolute value of $n$ to account for any eventuality of its signature. Using the substitution
$$
\psi=\Phi(z){\rm e}^{-z/2} z^l,
$$
we reduce (\ref{2.7}) to a degenerate hypergeometric equation
$$
z\frac{d^2\Phi(z)}{dz^2}+\left(\gamma-z\right)\frac{d\Phi(z)}{dz}-\alpha\Phi(z)=0,
$$
where
$$
\gamma=2l,\qquad \alpha=l-\frac{|n|}{\sqrt{2} \kappa \tau_{(2)PL}}.
$$

The degenerate hypergeometric function can be expanded into the series
$$
\Phi(z,\alpha,\gamma)=1+\frac{\alpha}{\gamma}\frac{z}{1!}+\frac{\alpha(\alpha+1)}{\gamma(\gamma+1)}\frac{z^2}{2!}+\frac{\alpha(\alpha+1)(\alpha+2)}{\gamma(\gamma+1)(\gamma+2)}\frac{z^3}{3!}+...
$$
which converges when $\alpha=-m$, where $m$ is an integer. It is this fact that allows us to retrieve the spectrum of (\ref{quli}) with $N=m+l$.

For example, consider a simple case when $l=1$, $m=0$. Here
$$
\psi(z)=z{\rm e}^{-z/2},\qquad u(z)=-\frac{4\kappa^2}{z},\qquad \kappa=\frac{|n|}{\sqrt{2}\tau_{(2)PL}},
$$
and the entire cosmological dynamics hinges on the signature of $n$. If $n>0$, the universe arises from the initial singularity with $z=0$ (here it is handy to use $z$ as a temporary variable), expands while $z\in (0,\,\,2)$, and then collapses in an asymptotical manner. The Hubble parameter and the acceleration ($a'=da/dz$) of this two-dimensional universe are:
\begin{equation}
H=\frac{2-z}{2nz}, \qquad \frac{a''}{a}=\frac{(z-z_-)(z-z_+)}{4n^2z^2},
\label{n>0}
\end{equation}
where $z_{\pm}=2\left(1\pm\sqrt{n}\right)$.

If $0<n<1$, we end up with a different dynamics. After initial singularity at $z=0$, the universe undergoes a period of accelerated expansion for $z\in (0,\, z_-)$, to be followed by a period of decelerated expansion during $z\in (z_-,\, 2)$. When the parameter $z$ reaches its critical value $z=2$ the expansion turns into a collapse: a decelerated one at first (when $z\in (2,\, z_-+)$), but the pace then picks up and for $z>z_+$ we have a phase of accelerated collapse which satisfies the condition
$$
\lim_{z\to+\infty}\frac{a''}{a}=\frac{1}{4n^2}.
$$

This type of dynamics will be mostly preserved if $n\ge1$ but with a caveat: such a universe will experience acceleration only once -- during the final pahse of collapse.

Using (\ref{2.1})--(\ref{2.4}) one can easily surmise the dynamics of the scalar field $\phi(t)$ and derive the exact form of the self-action potential $V(\phi)$. Indeed, the vacuum energy's density for the aforementioned values of $l$ and $m$ will be equal to
$$
\rho_{_{\Lambda}}=\frac{1}{8\pi G_{(2)}\tau_{(2)PL}^2},
$$
and since the collapse phase starts when $t_{_{STOP}}=\sqrt{2}n^{-1}\tau_{(2)PL}$, we conclude that the universe will {not} experience an immediate collapse only if $n\ll 1$. A simple integration yields:
\begin{equation}
\phi(t)=\pm\frac{\sqrt{2}}{4\sqrt{\pi G_{(2)} n}}\log\,t.
\label{phi}
\end{equation}

We shall choose the negative sign in (\ref{phi}) from physical considerations (this way the field decreases with time instead of increasing). It also proves useful to redefine the field variable as,
$$
-\phi\sim \chi=\log\,t,
$$
so that the self-action potential takes the following form:
\begin{equation}
V(\chi)=\frac{1}{16\pi G_{(2)}\tau_{(2)PL}^2n^2}\left((2-n)\tau_{(2)PL}^2{\rm e}^{2\chi}-2\sqrt{2}n\tau_{(2)PL}{\rm e}^{\chi}-n^2\right).
\label{poten}
\end{equation}

However, what about the case $n<0$? Here the picture changes drastically: the universe in this case arises not from an initial singularity but from the {Big Rip} singularity. This happens at $z=0$, then the universe collapses while $z\in (0,\, 2)$, until at $z=2$ it experiences a ``rebound'' (with a finite scale factor!). And then it undergoes a period of exponential expansion in a quasi-de Sitter regime. The corresponding formulas are rather cumbersome, so we will omit them here.

\section{A Fractal Horizon? }\label{sec:fractals}

In a recent article \cite{Bar-1} John Barrow has proposed a very interesting idea. According to him, while studying the various micrographs of a SARS-CoV-2 coronavirus he was struck by a sudden realization that an event horizon might theoretically have a fractal structure. {After all, we known that any type of interaction of black hole (BH) with its surroundings deforms the surface of its event horizon, including even the vacuum outside of a BH, which causes a slight decay in the radius of the event horizon, producing a famous Hawking radiation along the way. If we go all the way down from the macroscopic scale to the Plank scale we will see the BH surrounded by the spacetime foam} \cite{Wheeler,Misner}, {which might produce an interesting foam-like fractal (or quasi-fractal) perturbations in the event horizon of that BH} \cite{Bar-1}.
 Following this idea, Barrow considered a simplest generalization of a \linebreak Koch snowflake \cite{Koch}, where the standard spherical horizon of radius $r_0$ is extended by $M \ge 1$ additional ``spheres'', attached to it. However, this is just a first iteration. On a second iteration, every one of those ``spheres'' is further extended by $M$ more ``spheres'' of progressively smaller radius, etc (see \cite{video} for a 3D visualisation). By ``progressively smaller'' we mean that the radius of the new ``spheres''  on the $n$-th iteration will  be equal to $r_n=\lambda^n r_0$, where $\lambda<1$. After the the $n$-th iteration the total $d$-dimensional volume of the fractal horizon and its surface area will be equal to:
\begin{equation}
V_{(d)n}=\frac{\Omega_{d-1}r_0^d\left(1-\left(M\lambda^d\right)^{n+1}\right)}{d\left(1-M\lambda^d\right)},\qquad
A_{(d)n}=\frac{\Omega_{d-1}r_0^{d-1}\left(1-\left(M\lambda^{d-1}\right)^{n+1}\right)}{1-M\lambda^{d-1}}.
\label{4.0}
\end{equation}

{Note that, unlike the classical Koch snowflake which carries infinitely many ``spheres'' of progressively infinitesimal volume, within our framework there exists a natural lower limit on the size of new ``spheres'': the Plank volume (we are dealing with a quasi-fractal structure). In order to implement this limit, let us assume} that the volume of a smallest $d$-dimensional ``sphere'' is indeed equal to a $d$-dimensional Plank volume. i.e., that $n$ in (\ref{4.0}) satisfies the condition
\begin{equation}
n=\frac {\log (r_0/L_{(d)PL})}{\log 1/\lambda}.
\label{4.1}
\end{equation}

If we substitute (\ref{4.1}) into (\ref{4.0}) and use our Main Hypothesis that the $d$-dimensional surface area of the horizon with $r_0$ can only contain a whole number $N$ of $d$-dimensional Plank areas, we will get this:
\begin{equation}
\left(1-M\lambda^{d-1}\right)^{-1}\left(\frac{r_0}{L_{(d)PL}}\right)^{d-1}\left(1-M^{1+\log\left(\frac{r_0}{L_{(d)PL}}\right)/\log\left(\frac{1}{\lambda}\right)}\left(\frac{L_{(d)PL}\lambda}{r_0}\right)^{d-1}\right)=N.
\label{4.2}
\end{equation}

Note, that \eqref{4.2} explicitly depends on the product of two parameters: $M$ and $\lambda$. We can estimate this product using the following physical reasoning: if it is too small (i.e., $M\lambda \ll 1$) then the result will be essentially identical to the original, non-fractal formula (and if it is too small even the largest of those additional ``spheres'' will be smaller than the cut-off Plank's length). On the other hand, if $M\l \gg 1$ the ``spheres'' will begin to overlap, which is also unphysical. Thus, the product of $M$ and $\lambda$ shall be of magnitude $1$; following this train of thought, let us assume that $\lambda M=1$ (we can choose any other number for as long as it is of the same order of magnitude as $1$). Such assumption seriously simplifies \linebreak (\ref{4.2}), turning it into
\begin{equation}
\left(1-\frac{1}{M^{d-2}}\right)^{-1}
\left(\frac{r_0}{L_{(d)PL}}\right)^{d-1}
\left[1-\left(\frac{L_{(d)PL}}{Mr_0}\right)^{d-2}\right]=N.
\label{4.3}
\end{equation}

Denoting $r_0/L_{(d)PL}=x$, $M^{d-2} = \mu$ we can then rewrite (\ref{4.3}) as a relatively simple polynomial equation:
\begin{equation}
\mu x^{d-1}-x-N(\mu-1)=0.
\label{urra}
\end{equation}

If $\mu\gg 1$ (which happens when $\lambda \ll 1$), the solution of (\ref{urra}) can be nicely approximated by $x\sim N^{1/(d-1)}$, which essentially reproduces the old results (\ref{3.3}). For example, in three dimensions (\ref{urra}) will be a quadratic equation with a single positive root
$$
x=\frac{1+\sqrt{1+4N\mu(\mu-1)}}{2\mu},\qquad \lim_{\mu\to\infty} x=N^{1/2},
$$
while $d=4$ case produces a cubic equation with two complex conjugate roots (unimportant for us) and one real root:
$$
x=\frac{A^{1/3}}{6\mu }+\frac{2}{A^{1/3}}, \qquad A=12\mu \left[9N\mu (\mu -1)+\sqrt{3\mu \left(27N^2(\mu -1)^2\mu -4\right)}\right],
$$
where $\lim_{\mu\to\infty} x=N^{1/3}$. In other words, it appears that the procedure of ``quantization'' proposed in this paper remains stable even in the hypothetical case of fractal event horizons.

\section{In Conclusion: A Few More Words on the Koch Snowflakes}\label{sec:conclusion}

Before we wrap up, let us briefly discuss the results we have gained throughout our trek, and the avenues of research that remain
untrodden. We have seen that higher dimensional cosmological methods and models in more ways than one remain similar to the familiar three-dimensional models and methods. In particular, many analytic methods of integration originally designed for the three-dimensional Friedmann equations can be easily generalized on a $d$-dimensional case. This is easily illustrated in the Appendix \ref{sec:appendix} where we show that it is possible to take a general method, invented by Tipler in \cite{Tipler} to evaluate a lifetime of a closed three-dimensional universe filled by matter with a barotropic equation of state with $w$ = const, and then, by very slightly altering it, make the Tipler's method work for any arbitrarily dimensional universe; and even to get an exact and analytic expression for a total lifespan for a closed universe filled with a phantom energy.

Nevertheless, there do exist at least two points of divergence from the well-known results. First, our Main Hypothesis predicts that the universes with different numbers of dimensions will have different spectrums of the vacuum density \eqref{3.3}. We have seen that for $d=2$ the spectrum will be of a simple Coulomb type \eqref{3.5} with $E_{(2)} \approx 1/N^2$ (similar to a spectrum of a hydrogen atom). However, it is no longer true for higher dimensions: say, for $d=3$ the spectrum will be defined as $E_{(3)} \approx 1/N$, and for $d=4$ it will be $E_{(4)} \approx 1/N^{2/3}$. Thus, if the Main Hypothesis is correct, the number of dimensions in a universe is indelibly imprinted  in its possible values for the vacuum energy -- and, therefore, in the ultimate dynamics of that universe.

Another interesting point can be made about the limits on ``fractality'' of the horizons we have discussed in Section \ref{sec:fractals}. In that section we have discussed the Barrow's hypothesis of extending the horizons in a manner similar to the Koch snowlake \cite{Koch}. The Koch snowflake, like its counterparts Sierpi\'nski Gasket  \cite{Serp} and the Menger Sponge \cite{Menger}, has the following fascinating property: while its volume is finite, its surface area might be infinite (in case of the proper, two-dimensional, Koch snowflake, it means a finite surface with an infinite perimeter). In Section \ref{sec:fractals} we have restricted ourselves to the fractal horizons of finite $d$-volumes and finite surface areas. However, if the numbers $M$ and $\l$ in \eqref{4.0} happen to satisfy the double inequality
\begin{equation}
\frac{1}{\lambda^{d-1}}<M<\frac{1}{\lambda^d},
\label{Koch}
\end{equation}
we will instead have a universe of a finite volume (contained within the horizon), but {infinite} $d$-dimensional surface area of the horizon. And, since it is the latter that defines the Bekenstein bound \cite{Bek-1,Bek-2}, the condition (\ref{Koch}) actually leads to an infinite entropy.

Granted, since an infinite entropy is tantamount to an infinite ``concealed'' information, this begs a question: how can this have any physical meaning? In fact, it is most probable that in three-dimensions it most certainly does not. However, it is no longer so certain in the case of higher dimensions! In order to understand why that is, let us recall a recent \linebreak paper~\cite{Bar-3} by Barrow, where he has demonstrated that the event horizons of the \linebreak $d$-dimensional black holes cease to effectively shield the central singularities when $d>3$. In other words, in the 4- and higher-dimensional universes the cosmic censorship can be violated, letting loose the monstrous naked singularities \cite{Bar-3}. The reader will agree with us on the harsh epithet ``monstrous'', once she recalls that near the singularity all local physical laws must crumble. Of course, it ought to be said that most scientists stick to a view that all classical singularities will eventually be entirely undermined by the quantum effects or, at the very least, will get localized as some exotic points with extreme yet finite values of the scalar curvature, energy-momentum tensor etc. Unfortunately, it is a position to which a very simple counter-argument exists: the absence of singularities as space-time boundaries \cite{Large-Str} has an unfortunate side-effect of cosmological action becoming \linebreak singular \cite{Bar-4,Bar-Tip}. And this is an even more disastrous phenomena, since it renders any attempt to calculate a, say, Feynman integral moot and meaningless.

Subsequently, Barrow and Tipler have proposed an idea that the cosmological singularities, as well as those hidden inside of the black holes, all have to be treated as a ``lesser evil'', dangerous but ultimately benevolent as it protects us from a much greater ``evil'': a singular action. If their position is correct, then we should not expect the long-awaited ``quantum gravity'' theory to help us get rid of cosmological singularities (for example, via some kind of ``rebound'' etc.). On the other hand, in order to accept the Barrow and Tipler position we must know for certain that there does exist some version of ``cosmic censorship'' which shrouds the singularities from the outside observers. All this makes the mere possibility of a ``nakedness'' for singularities in higher dimensions, announced by Barrow in \cite{Bar-3}, all the more devastating. And so, we have arrived at an impasse.

There might be two possible solutions to our predicament. The first one would be to do away with the concept of higher dimensions. Of course, this also means dispensing with the string theory (at least in its current form). This is hardly an improvement! A second way, while not being so radical is still rather exotic: according to it, we shall assume that the surface area of a cosmological event horizon at $d>3$ must be infinitely large. The reasoning behind this assumption is actually straightforward; after all, a finite surface area leads to a finite entropy and, thus, to a finite number of possible microstates within. Any such theory will fold under the infinite pressure of singularity. If, however, the entropy is allowed to grow indefinitely, we will get infinitely many distinct possible microstates. And then we at least get some fighting chance to perhaps associate some of them with the divergent processes that accompany the singularity. Granted, what we have adumbrated here is at this stage merely a hypothesis, and speculative at that; but it at least gives some glimmer of hope in our struggle with a seemingly unsurmountable problem noticed by Barrow in \cite{Bar-3}.

\appendix \section{How to Study a \boldmath{$d$}-Dimensional Friedmann Universe with an Arbitrary Barotropic Equation of State}\label{sec:appendix}

In \cite{Tipler} a very elegant method has been proposed by Frank Tipler, which method allows one to easily integrate the Friedmann equations for a closed universe if it is filled with matter whose equation of state is
\begin{equation}
\frac{p}{c^2}=\left(\gamma-1\right)\rho,
\label{T1}
\end{equation}
and where $\gamma={\rm const}$ is an arbitrary adiabatic index. In fact, the method is so robust that it can be applied to both the closed and the open universes (the flat case is anyways trivial). However, Tipler was mostly interested in the closed case, because his method has allowed him to analytically calculate the total lifetime of such a universe, from the initial singularity to the final one. And, as we shall see below, the Tipler's method remains quite applicable even in the $d$-dimensional case.

We will consider two examples: a close $d$-dimensional Friedmann universe whose constituent matter satisfy certain energy conditions (see below), and a similarly closed universe filled with a phantom field. As we shall see, in both of these examples the life of the universe is finite and is sandwiched between two singularities. Of particular interest here will be the phantom example (with $\gamma<0$, $\gamma={\rm const}$), since its boundary singularities will be none other then the Big Rips, which in and of itself appears to be very unusual (see, however, the previous article\cite{PedroTema}, where this type of solutions has been first procured, albeit stemming from a more complex equation of state).

Let us integrate the continuity equation, taking into account (\ref{T1}), using the density formula from (\ref{2.1}). Temporarily abstaining from choosing the curvature $k$ of the universe, we get:
\begin{equation}
\frac{\dot a^2}{a^2}=\frac{16\pi G_{(d)}R^2 a^{-\gamma d}}{d(d-1)}-\frac{kc^2}{a^2},
\label{T2}
\end{equation}
where $R^2$ is the integration constant and $c$ is the speed of light.

Next, let's introduce a new variable: $a=y^{2/(\gamma d-2)}$ and switch to a new, {conformal time} $\tau$, which is related with the standard time $t$ via the formula
\beqn
dt = a(\tau) d\tau
\enqn

For brevity, we will from now on denote the derivative with respect to $\tau$ with an apostrophe, so that ${\dot a}a=a'$. It is then straightforward to show that with a new variable the Equation (\ref{T2}) takes a form:
\begin{equation}
y'^2+k\left(\frac{c(\gamma d-2)}{2}\right)^2y^2=\frac{\pi G_{(d)}}{d(d-1)}\left(2R(\gamma d-2)\right)^2.
\label{T3}
\end{equation}

For a closed universe with $k=+1$ this equation is an integral of the energy of a harmonic oscillator, which oscillates with a frequency $\omega=c|\gamma d-2|/2$. Hence, it is easy to derive the shapes of $y(\tau)$ and $a(t)$ in the parametric form:
\begin{equation}
a(\tau)=\left(\frac{4R}{c}\sqrt{\frac{\pi G_{(d)}}{d(d-1)}}\right)^{2/(\gamma d-2)}\left(\sin \omega\tau\right)^{2/(\gamma d-2)},\qquad t(\tau)=\int_0^{\tau} d\eta a(\eta).
\label{T4}
\end{equation}

If $\gamma>0$, the fields of matter satisfy the weak energy condition. Let us strengthen this a bit further by choosing $\gamma>2/d$. At this point we should also point out that for the cold matter ($\gamma=1$) this condition will be satisfied when $d>2$, while for the radiation-filled universe (with $\gamma=(1+d)/d$) the condition lowers to $d>1$. In other words, the condition we have just imposed is a very mild one. In any way, (\ref{T4}) with condition will describe a universe which begins at $\tau=0$, expands until the scale factor reaches it maximal value
\begin{equation}
a_{\rm{max}}=a(\tau=\frac{\pi}{2\omega})=\left(\frac{4R}{c}\right)^{2/(\gamma d-2)}\left(\frac{\pi G_{(d)}}{d(d-1)}\right)^{(\gamma d-2)^{-1}},
\label{T5}
\end{equation}
and then at $\tau=\pi/\omega$ a collapse commences. Taking the integral in (\ref{T4}) within the specified limits, we will end with a total lifetime $T_{\gamma,(d)}$ of this $d$-dimensional universe:
\begin{equation}
T_{\gamma,(d)}=\frac{2\sqrt{\pi}}{c}\left(\frac{4R}{c}\right)^{2(\gamma d-2)^{-1}}\left[\frac{\pi G_{(d)}}{d(d-1)}\right]^{(\gamma d-2)^{-1}} \frac{\Gamma\left(\frac{\gamma d}{2(\gamma d-2)}\right)}{\Gamma\left(\frac{1}{\gamma d-2}\right)}.
\label{T6}
\end{equation}

It is easy to see that for $d=3$, the Equation (\ref{T6}) is similar to the Tipler's formula he derived in \cite{Tipler}. Another notable result can gleamed by comparing the lifespans of different universes with the same $\gamma$ and $a_{\rm{max}}$ but different $d$; it appears that the bigger $d$ is, the shorter the lifespan gets. For example, here is how long the universe, dominated by the radiation, will exist:
\begin{equation}
T_{1+1/d,(d)}=\frac{2 a_{\rm{max}}\sqrt{\pi}}{c(d-1)}\frac{\Gamma\left(\frac{d+1}{2(d-1)}\right)}{\Gamma\left(\frac{d}{d-1}\right)}.
\label{T7}
\end{equation}
If $d=3$ the relationship (\ref{T7}) gives a well-known prediction: $T_{1+1/3, (3)}=2 a_{\rm{max}}/c$. This timespan will get progressively smaller the bigger $d$ gets; say, for $d=4$, $d=5$ and $d=10$ the lifespan will be equal to, subsequently, $1.49 a_{\rm{max}}/c$, $1.2 a_{\rm{max}}/c$ and $0.61 a_{\rm{max}}/c$.

Alternatively, if the unverse is filled with the cold matter (i.e., $\gamma=1$),~the \linebreak Formula (\ref{T6}) predicts that
\begin{equation}
T_{1,(d)}=\frac{2 a_{\rm{max}}\sqrt{\pi}}{c(d-2)}\frac{\Gamma\left(\frac{d}{2(d-2)}\right)}{\Gamma\left(\frac{d-1}{d-2}\right)},
\label{T8}
\end{equation}
and if, as before, we choose $d=3,\,4,\,5,\,10$, we will get the following lifespans: $\pi a_{\rm{max}}/c$, $2 a_{\rm{max}}/c$, $1.5 a_{\rm{max}}/c$ and $0.68 a_{\rm{max}}/c$.

In other words, we see that the lifetime of a closed universe decreases with the addition of new dimensions. Why should this be so? In order to understand it, we have to recall an important fact about the critical density of a $d$-dimensional universe. According to (\ref{T2}), it is proportional to a number of dimensions squared (it is proportional to $d(d-1)$, to be precise), so by adding new dimensions we effectively pump up the critical density. On the other hand, the density of a closed universe must by definition exceed the critical density, or it will not be closed at all. And, of course, the bigger the density gets, the sooner the closed universe stops expanding and begins to collapse, severely reducing its own lifespan in the process.

Finally, let us take a look at a more exotic case of phantom fields with $\gamma<0$. We have already performed all the preliminary calculations, so we can jump straight to the results. In the conformal time the Friedmann solution will look like
\begin{equation}
a(\tau)=a_{\rm{min}}\left(\sin \left(\frac{c(|\gamma| d+2)\tau}{2}\right)\right)^{-2/(|\gamma| d+2)}
\qquad a_{\rm{min}}=\left(\frac{c}{4R}\sqrt{\frac{d(d-1)}{\pi G_{(d)}}}\right)^{2/(|\gamma| d+2)},
\label{T9}
\end{equation}

and we have a universe that coalesces from the Big Rip singularity from $\tau=0$, collapses to its minimal size $a=a_{\rm{min}}$ on the interval $\tau=\pi/c(|\gamma| d+2)$, and then undergoes a rapid expansion which concludes in a second, and final, Big Rip singularity. A total lifespan in the normal time can, as usual, be calculated via the integration and is equal to
\begin{equation}
T^{\rm{phantom}}_{\gamma,(d)}=\frac{2 a_{\rm{min}}\sqrt{\pi}}{c(|\gamma| d+2)}\frac{\Gamma\left(\frac{|\gamma| d}{2(|\gamma| d+2)}\right)} {\Gamma\left(\frac{|\gamma| d+1}{|\gamma| d+2}\right)}.
\label{T0}
\end{equation}

In an interesting article \cite{Omo} an interesting problem has been studied: a possibility for a universe to alter its topology while experiencing the Big Rip singularity. Before we conclude this article we would like to point out that we this problem might lead to even more curious results in the $d$-dimensional cosmologies. After all, more dimensions usually implies more complex topologies, so it would be only natural to expect some rather peculiar scenarios there.

\section*{Acknowledgments} \label{sec:Acknowledgements}
The work was supported by the programm 5-100 of I. Kant Baltic Federal University.

\end{document}